\documentclass{rspublic}
\usepackage{pslatex,graphicx,dcolumn,bm,natbib,subfigure}

\begin{document}

\title[Dynamics in a Stress Landscape]{Rheology of Granular Materials: Ê Dynamics in a Stress Landscape}

\author[D. Bi, B. Chakraborty]{Dapeng Bi, Bulbul Chakraborty}

\affiliation{Department of Physics, Brandeis University,
Waltham, MA 02454, USA}

\label{firstpage}

\maketitle

\begin{abstract}{granular,rheology,stress}
We present a framework for analyzing the rheology of  dense driven granular materials, based on a recent proposal of  a stress-based
ensemble.  In this ensemble fluctuations in a granular
system near jamming are controlled by a temperature-like parameter, the angoricity, which is conjugate to the stress of
the system.  In this paper, we develop a model for slowly driven granular materials based on the stress
ensemble and the idea of a landscape in stress space.  The idea of an activated process driven by the angoricity has
been shown by Behringer {\it et al} (2008) to describe the logarithmic strengthening of granular materials .  Just as in
the Soft Glassy Rheology (SGR) picture, our model represents the evolution of a small patch of
granular material (a mesoscopic region) in a stress-based trap landscape. The angoricity plays the role of the
fluctuation temperature in SGR.   We determine (a) the constitutive equation, (b)
the yield stress, and (c) the distribution of stress dissipated during granular shearing experiments, and compare these
predictions to experiments of Hartley \& Behringer (2003).  

\end{abstract}

\section{Introduction}

A striking feature of dry granular materials and other athermal
systems is that they form force chain networks, in which large forces
are distributed inhomogeneously into linear chain-like structures, in
response to applied stress (Majmudar \& Behringer (2005)).    A number of
recent experimental studies have visualized and quantified force chain
networks using carbon paper (Jaeger {\it et al} (1996)) and photo-elastic
techniques (Howell \& Behringer (1999), Veje {\it et al} (1999)).  These studies have demonstrated that the characteristics of 
force chain networks are acutely sensitive to the nature of the prepared state, especially near the jamming
transition (Majmudar \& Behringer (2005)).  For example, in isotropically compressed systems,
force chain networks are ramified with only short-ranged spatial
correlations of the stress.  In contrast, in sheared systems, aligned
force chains give rise to long-ranged spatial correlations of
the stress in the direction of the shear.  Any theory of granular rheology has to, therefore, incorporate the effects of
these grain-level structures on the macroscopic, collective response.  

It has been long realized that classical elasticity theory or Newtonian fluid dynamics is inadequate for describing the 
response of granular materials for a couple of  important
reasons (Bouchaud (2003), Cates {\it et al} (1998), Blumenfeld (2004)): (1) tensile stresses are completely absent in dry granular materials,
and therefore the cohesion of the granular assembly is induced by the applied stress making the zero-stress state
ill defined, (2) there is an indeterminacy of the forces at the microscopic level due to friction and disorder, (3) granular materials are
athermal and dissipative, therefore
there is no established statistical framework that bridges the structure  at grain  scales
to a continuum elasto-plastic theory at large length scales.  In addition, granular solids 
often occur in isostatic states in which the
number of degrees of freedom matches the number of constraints (Tkachanko \& Witten (1999), Moukarzel  (1998)).  It
has been shown that the critical properties of the isostatic point,
not elasticity theory, determine the mechanical response of these
marginal solids (Wyart (2005)). 

A framework, known as soft-glassy rheology has been used widely to understand the rheology of materials such as foams,
colloids, and grains (Sollich (1998)).  In this work, we use a recently proposed statistical framework, the stress
ensemble for describing the response of static granular media (Henkes \& Chakraborty (2009), Henkes {\it et al} (2007)), to formulate a theory of
the response of slowly (quasistatic) sheared granular media and discuss its predictions for granular rheology.

\section{Stress Ensemble}
As shown
previously (Henkes \& Chakraborty (2009), Ball \& Blumenfeld (2002)), the stress field of a
a mechanically stable  granular material can be fully described by a spatially-varying
scalar field $\psi$ in two dimensions, and a tensorial field in higher dimensions (Henkes \& Chakraborty (2009)). 
These fields can be used to establish a rigorous conservation
principle, valid for grain assemblies in mechanical equilibrium, at the grain level in two
dimensions and at a continuum level in higher dimensions (Henkes \& Chakraborty (2009)).  
This conservation principle, in conjunction with a maximum entropy hypothesis (Jaynes (1957)), leads to a generalization
of equilibrium thermodynamics to the ensemble of mechanically stable granular states (Henkes \& Chakraborty (2009), Henkes {\it et al} (2007))
where the new conserved 
tensorial quantity,
 $\hat \Sigma$ to be discussed below,  plays the role of energy. 
The microcanonical version of this ensemble is characterized by the complexity, a measure of the number of grain
configurations compatible with a given value of $\hat \Sigma$.  The assumption of entropy maximization leads to the
definition of a temperature-like intensive variable, the angoricity (Henkes \& Chakraborty (2009), Blumenfeld \& Edwards (2003)), which is a
tensor.  The predictions of this stress ensemble have been remarkably successful in describing results of
simulations (Henkes {\it et al} (2007), Lois {\it et al} (2009))
and experiments in both static (Majmudar \& Behringer (2007), Lois {\it et al} (2009)) and slowly driven granular media (Behringer {\it et al} (2008)).  
As an example, the stress ensemble framework and the constraint that all contact forces are non-negative has been used
to develop a Ginzburg-Landau type action for jammed granular systems in terms of $\psi$.  The theory has been used to
predict the spatial correlations
of stress in systems subjected to isotropic compression (Henkes \& Chakraborty (2009)), and shear (Lois {\it et al} (2009)), and to construct a
mean-field theory of unjamming under isotropic compression (Henkes \& Chakraborty (2009)).  In slowly sheared granular systems, the
observation of a logarithmic strengthening has been explained by the stress ensemble approach(Behringer {\it et al} (2008)).  In this
paper, we present a detailed development of  the theory of slowly sheared granular matter, and make falsifiable
predictions.  Preliminary tests of the theory, based on comparison to experiments using photo-elastic
disks (Hartley \& Behringer (2003)) are also presented.

\subsection{Conservation principle and Complexity} To understand the mechanical response of
granular materials, one needs a theoretical approach that can bridge
the gap between microscopic, grain-level quantities and macroscopic,
collective properties.  Fluctuations are inherently related to the number of microscopic states available under a
given set of macroscopic parameters.  In equilibrium thermodynamics, the microcanonical entropy, or its derivatives in
other ensembles,  is the measure we use to calculate fluctuations and response.  Conservation of energy allows for a
rigorous definition of this measure since the states with different energies are not mixed by dynamics.  In
disordered systems such as spin glasses,  the concept of complexity has been useful in formulating a framework for
understanding collective properties (Bouchaud {\it et al} (1996)).  Complexity is a measure of the number of
states associated with a free energy minimum. and has proven to be a useful concept for disordered systems that have many
metastable states.  In mean-field models, these minima are separated by barriers that
diverge in the thermodynamic limit, and one can in principle count the number of states unambiguously.  Can we 
identify a physical variable in granular materials, which also have many metastable states (Bouchaud (2003)), which is
conserved by natural dynamics and, therefore, leads
to a formalism for defining complexity?  The lack of a natural dynamics in granular systems makes this a difficult
proposition.   In mechanically stable states, however, there is a topological conservation law that allows us to
proceed to define the analog of complexity.  The topological nature implies that a change in this physical variable
can be achieved only through rearrangements that involve the boundaries or the whole system.  A gedanken experiment
serves to illustrate the conservation law.  If we draw an imaginary line through a grain assembly and calculate the
value of the 
total normal force being transmitted along this line, then force balance ensures that this value remains unchanged as
this imaginary line is translated perpendicular to itself across the assembly (Metzger (2008)).  A formal theoretical
framework can be formulated using the force-moment tensor $\hat \Sigma =
\int d^dr \hat \sigma(\mathbf{r})$, where $\hat
\sigma(\mathbf{r})$ is the local stress tensor. It has been shown (Henkes \& Chakraborty (2009), Henkes {\it et al}
(2007))  that  $\hat \Sigma$ 
depends only on the boundary conditions of the packing. 
Thus,  the phase space of all mechanically stable configurations can be divided into sectors labeled by  $\hat
\Sigma$. Configurations in different sectors are disconnected under any local dynamics, and $\hat \Sigma$ plays a
role similar to total energy in equilibrium thermodynamics or free-energy minima in spin glasses.  Based on some very
general principles, such as the factorization of states (Bertin {\it et al} (2006)), the
conservation of $\hat
\Sigma$ has
been shown to lead to an intensive variable, $\hat \alpha = {\partial S(\hat \Sigma) \over \partial  \hat \Sigma}$, where $S(\hat
\Sigma)$ is the entropy of the sector labeled by $\hat \Sigma$ (Henkes \& Chakraborty (2009)). The
inverse of $\hat
\alpha$, the
angoricity (Edwards \& Blumenfeld (2007)) is expected to play the same role as temperature in equilibrium thermodynamics,
if processes that create granular assemblies achieve entropy maximization. 
Comparisons with simulations have demonstrated that a mechanically stable assembly of grains has the same value of
angoricity throughout its interior, while $\hat \sigma ({\bf r})$, the local stress fluctuates.  The probability of
occurrence of a microscopic state $\nu$ within a mechanically stable grain packing, maintained at an angoricity, 
$\alpha$,   is: 
\begin{equation}
P_{\nu} = (1/Z)e^{-\hat \alpha : \hat \Sigma_{\nu}} ~,
\label{Boltzmann}
\end{equation} 
analogous to the Boltzmann distribution (Henkes \& Chakraborty (2009)).  The state $\nu \equiv \lbrace \bf{
r_{i}}, \bf {f_{ij}} \rbrace$, where $\bf{r_{i}}$ denote the positions of the grains, and $\bf{f_{ij}}$ the set of 
contact forces.

A natural
question to ask is whether the similarities between temperature and angoricity extend to dynamics of slowly deformed
granular systems, situations analogous to dynamics of thermal systems close to thermal equilibrium where stochastic
dynamics such as relaxational,  Langevin dynamics provides a good description (Goldenfeld (1992)).
In the granular
context, we are restricting our attention to systems that are close to granular equilibrium defined by Eq.
\ref{Boltzmann}.  As in systems close to thermal equilibrium, we can construct a stochastic, Markov process based on the
concept of detailed balance:
\begin{equation}
{W_{\nu \rightarrow \nu^{\prime}} \over W_{\nu^{\prime} \rightarrow \nu}} = {P_{\nu^{\prime} } \over P_{\nu}}
\label{detailedbal}
\end{equation}
This is an additional assumption beyond the ones entering the construction of the stress ensemble that leads to the
distribution of mechanically stable states, $P_{\nu}$. 
Stochastic equations, based on detailed-balance conditions lead to the Boltzmann-like distribution, Eq. \ref{Boltzmann}, as the time-independent distribution of states (Goldenfeld (1992)).    If, however, the stress-landscape (or energy landscape in a thermal ensemble) is such that complete equilibration is not possible then the stochastic equations lead to a  rate of escape over a barrier $\Delta \hat \Sigma$, which  is the well-known Kramers rate:
$e^{-\hat \alpha : \Delta \hat \Sigma}$ (Kramers (1940)).  A simple example of a landscape where equilibration is not possible is one that has a single local minimum and a single local maximum.  In that situation the system ``escapes'' over the barrier with a rate prescribed by Kramers but cannot equilibrate.  As will become clear from the discussion below, trap models describe systems that escape over barriers with the Kramers rate, however, subsequent sampling of traps is described by a quenched distribution and not by detailed balance.
We adopt this form of activated dynamics to analyze the rheology of dense, quasistatic granular flows.

Our framework of granular rheology is, therefore, based on a rigorous conservation
principle and postulates, similar to the ones adopted in the development of equilibrium
thermodynamics (Callen (1957)).  These postulates are (a) factorisability of the dynamics-dependent
frequency with
which each grain configuration is accessed (Henkes \& Chakraborty (2009)), which allows for the
definition of the angoricity,
(b) maximization of entropy, which implies equality of the angoricity across a system in granular equilibrium and a
Boltzmann-like distribution (Eq. \ref {Boltzmann}) (Henkes \& Chakraborty (2009)),
and (c) satisfaction of detailed balance  by the rates of transition between microscopic states.

\section{Trap models and rheology}
As discussed earlier, the 
existence of a large number of different microscopic metastable states that are macroscopically equivalent
puts granular materials into a wider class of previously well studied systems, including gels, glasses, colloids,
emulsions, polymers and foams. A minimal model that encapsulates these features, and leads
to a glass transition is the ``trap model" (Monthus \& Bouchaud (1996)).
In this model, one
considers a system made of independent subsystems of a certain size $\xi $, where each subsystem acts self-coherently
and independent of the others. Their dynamics involve hopping between different metastable states aided by some kind of
fluctuation. This idea of a random walk in a rugged landscape has its roots in the context of glasses (Bouchaud \& Georges  (1990)). 
Below, we discuss the trap model and its generalization, which is the framework
of soft glassy rheology (SGR) (Sollich (1998)) in some detail.

\subsection{Bouchaud's trap model for glasses}  
Monthus \& Bouchaud (1996)  constructed a one-element model for glasses. In this
model, there exists an
energy landscape of traps with various depths $E$. An element hops between traps when activated, where the fluctuation is assumed to be
thermal.  At a temperature $k_{B}T \equiv 1/\beta$, the probability of being in a trap with depth $E$ at time
$t$ 
evolves according to:
\begin{equation}
\frac{
\partial }{
\partial  t}P (E, t)=-\omega _0 e^{-\beta  E}P (E,t)+\omega (t) \rho (E)
\label{bouchaudEOM}
\end{equation}
where,
\begin{equation}
\omega (t)=\omega _0\left\langle e^{-\beta  E}\right\rangle {}_{P(E, t)}=\omega _0\int dE P (E,t) e^{-\beta  E}
\label{bouchaudomega}
\end{equation}
The first term on the right-hand-side (rhs) of Eq. \ref{bouchaudEOM} is the rate of hopping out of a trap, where
$e^{-\beta E}$ is the activation factor and $\omega_0$ a frequency constant. The model assumes that
choosing a new trap is independent of history, so that a new trap is chosen from a 
distribution of trap depths $\rho$(E) that reflects the underlying disorder in the glassy models. Eq. \ref{bouchaudomega}
gives the average hopping rate. 
This rate, 
multiplied by the distribution of trap depth, Eq. \ref{bouchaudEOM} gives the probability rate of choosing a new trap. 

The existence of a glass transition in this model can be demonstrated as follows: assume the distribution function has a
simple exponential tail such as:
\begin{equation}
\rho (E) \propto e^{-\beta _0 E}~,
\label{bouchaudDOS}
\end{equation}
where $\beta _0$ is a fixed parameter describing the disorder in the inherent energy landscape. The physical
justification for having an exponentially decaying $\rho (E)$ is borrowed from systems with quenched random disorder,
such as spin glasses, which use extreme value statistics (Monthus \& Bouchaud (1996)) (We will later justify applying the
same form of distribution to granular materials). Then we can easily solve for the steady-state ($\frac{\partial }{\partial  t}P (E,
t)=0$) solution: 
\begin{equation}
P_{\text{eq}}(E)\propto e^{\beta  E}\rho (E) \propto e^{\left(\beta _0-\beta \right)E}
\label{bouchaudSOLN}
\end{equation}
Immediately, we see that Eq. \ref{bouchaudSOLN} is non-normalizable for $\beta >\beta _0$ or $T<T_0$. This shows that
below a temperature $T_0$, the system is out of equilibrium; it is non-ergodic and ages by evolving into deeper and
deeper traps. Therefore, we call $T=T_0$ a point of glass transition. (Monthus \& Bouchaud (1996)). 

It should be noted that dynamical equations of the trap model do not lead to a Boltzmann distribution at $T$ since the equations do not obey detailed balance, and specifically,  the traps are sampled from a quenched distribution.   The escape rate from a trap is, however, determined by the Kramers process.

\subsection{Soft Glassy Rheology}

In (Sollich (1998)), a simple phenomenological model termed ``Soft Glassy Rheology" (or SGR) was proposed to describe
the anomalous rheological properties of ``soft glassy materials". The model incorporated Bouchaud's glass trap model and an
extra degree of freedom, the strain. 
The model assumes a macroscopic soft glassy material can be sub-divided into a large number of mesoscopic regions, each
having a linear size $\xi$. With each mesoscopic region one can then associate an energy $E$ and a strain variable $l$,
which both evolve with time. The number of mesoscopic regions must be made large enough so that ensemble averaging can
be performed over them to yield macroscopic properties of soft glassy materials. Similar to Bouchaud's trap model, the
SGR model assumes the mesoscopic regions ``live" in an energy landscape. This is a mean-field energy landscape in that it
is not characterized by a metric,  and is defined only by the distribution $\rho (E)$. The new strain variable describes
local elastic deformation of the mesoscopic regions, so that $l$ contributes quadratically to the energy $E$. Since a
strain variable was added to the SGR, the model can now describe material under imposed shear strain, also in a
mean-field spirit, all mesoscopic regions respond uniformly to externally imposed shear strain. Different from
Bouchaud's trap model, thermal fluctuations are unimportant in soft materials ($k_B T$ is too small to cause structural
rearrangements). In the SGR model, it is the fluctuations in the elastic energy that facilitate structural
rearrangements. This fluctuation is determined by a temperature-like quantity $x$ called the ``noise level". As a result,
the
escape rate from a trap becomes $e^{-E / x}$. The use of $x$ is a mean-field approach to
describing all interactions between mesoscopic regions.

\subsection{Incorporating the stress ensemble into the SGR framework} 
The idea of using a temperature-like quantity to replace thermal fluctuations in 
the SGR model is reminiscent of the stress ensemble for granular materials, where 
the inverse angoricity, $1/ \hat \alpha$, plays the role of the noise, $x$,  in SGR.  The activated process in the stress
ensemble is a result of a coupling between different mesoscopic regions through stress fluctuations.  Each mesoscopic region can be viewed as existing in a
bath of stress fluctuations, which is characterized by the angoricity, and these fluctuations can lead to activated processes
analogous to those occurring in thermal systems. Since the granular material in relevant experiments (Hartley\&
Behringer (2003)) are sheared setups, and to simplify the model, we will use a scalar model that incorporates only the shear components of the 
stress through $\Gamma$, the deviatoric part of $\hat \Sigma$, and 
$\alpha$,  the shear component of the inverse-angoricity tensor (referred to as shear-angoricity below), 
which represents the bath of stress fluctuations that the mesoscopic region is in contact with.


To adopt the SGR framework to the stress ensemble, we need to define metastability and escape processes in stress  rather
than energy space. Since the stress ensemble is based on the premise that there
are many states $\lbrace {\bf r_{i}}, {\bf f_{ij}} \rbrace$ with a given $\Gamma$, and that these states are broadly
sampled even if not with equal weights (Henkes, 2009),  we assume that mescoscopic regions of granular assemblies can be in metastable
equilibrium characterized by the Boltzmann-like distribution (Eq. \ref{Boltzmann}).   A dynamics obeying detailed balance  (Eq. \ref{detailedbal}) would lead to this distribution.
As mentioned earlier, the trap model and SGR, however, describe systems in the presence of quenched disorder which define the distribution barrier heights (or equivalently trap depths) $\Delta \Gamma$ that can be crossed in a finite time.   The system can escape from a metastable state represented by a trap.  Stochastic equations based on the stress ensemble allows us to adopt the Kramers' escape rate approach.  Replacing temperature by the shear angoricity,  the escape rate is therefore, $e^{-\alpha \Delta
\Gamma}$, which replaces $e^{-\beta E}$ of Bouchaud's trap model.
As in the original trap model
framework,  the exploration of the traps in stress space, following this escape is assumed to be determined by the
intrinsic, quenched-in distribution of
trap depths, and this dynamics does not lead to the Boltzmann-like distribution, Eq. \ref{Boltzmann} in the long-time limit.   The quenched-in distribution of trap depths determines whether complete equilibrium is attainable  in
the absence of shearing (Monthus \& Bouchaud (1996)).  
The activated process in the original trap model was thermal, and in SGR the activation is controlled by the noise
$x$.  In the scalar version of the stress-ensemble, the activation is in stress space and is controlled by the
angoricity $\alpha$.  
The disorder in grain packings is  represented by an intrinsic distribution of trap depths, just as in SGR or the
original trap model.

We
can now write a generalization of the SGR model to granular materials using the stress ensemble:
\begin{equation}
\frac{d}{d t}P \left(\Gamma _m,\Gamma , t\right)=-\omega _0e^{-\alpha  \left(\Gamma _m-\Gamma \right)}P \left(\Gamma
_m,\Gamma , t\right)+\omega (t) \rho \left(\Gamma _m\right) \delta (\Gamma )
\label{sgrEOM}
\end{equation}
Eq. \ref{sgrEOM} describes the  evolution of the probability of finding a mesoscopic region with a "yield stress"
$\Gamma_{m}$ (again, strictly $\Gamma_m$ has the unit
of force moment) and an instantaneous stress $\Gamma$.   The yield stress $\Gamma_m$ is
used to define the depth of the 
trap the system is in, and since $\Gamma$ is the instantaneous stress of the system, the barrier
is $\Delta \Gamma = \Gamma_{m} - \Gamma$.  The trap depth $E$   of SGR or the trap model is therefore replaced in the
stress ensemble by $\Gamma_{m}$.

\begin{figure}[htp]
\centering
\includegraphics[width=4in]{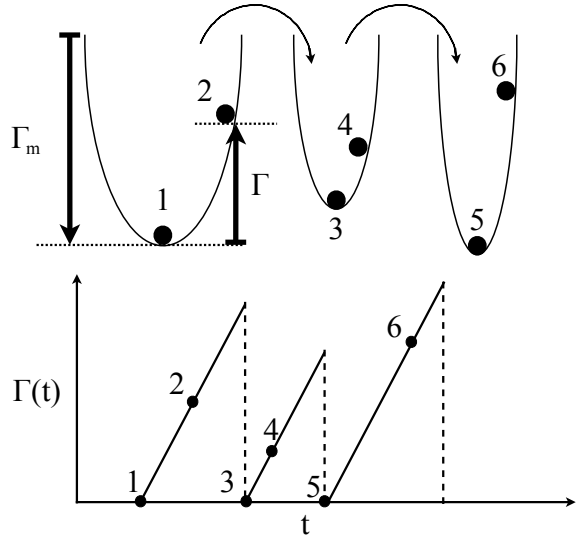}
\caption{Schematic showing dynamics of the model in the stress landscape. The black dots with the numbering (1 thru 6) 
represent the state of the mesoscopic region at different times. In the top figure, each  ``trap" in the landscape is
given by its depth $\Gamma_m$. The amount of stress buildup, while still in the trap, is indicated by the height $\Gamma$ from the
bottom of the trap. For ``1", the mesoscopic region has just fallen into the trap and it has no stress build up, 
so it has value of $\Gamma=0$ in the bottom figure. As constant shear stress is applied to the mesoscopic region, its
stress starts to build up. In the top figure, ``2" is at a height from the bottom of the trap; and in the bottom figure,
``2" has increased linearly in value from ``1". Next, via the activated process, the mesoscopic region makes a jump from
one trap to another. Note the traps shown are intentionally made disjoint, this is to indicate that every trap can be
accessed from every other trap and there is no specific connectivity.} 
\label{fig:landscape}
\end{figure}

The model accounts for the successive  stress buildup and yielding events inside a mesoscopic region.
When the instantaneous stress (due to external driving) is less than the yield stress, or $\Gamma < \Gamma_m$, the
mesoscopic region
is considered to be inside a ``trap" of depth $\Gamma_m$. It is possible to yield below the yield stress because of the
activated nature of the process. At this point the mesoscopic region ``sees" a reduced trap depth (i.e. effective
barrier height): $\Gamma_m - \Gamma$, and hence can undergo an activated process to ``hop" out of the trap with a rate
proportional to $e^{-\alpha  \left(\Gamma _m-\Gamma \right)}$. 
Another process through which the mesoscopic regions can yield is that if no activated process takes place while in the
trap ($\Gamma < \Gamma_m$), $\Gamma$ can increase due to external applied shear up to the point $\Gamma = \Gamma_m$
where the mesoscopic region yields, hence leaving the trap.  
This is the only possible process at the limit of zero
angoricity ($1/\alpha \to 0$), and is not explicitly included in Eq. \ref{sgrEOM}. Either way, after the mesoscopic region
leaves a trap, it choose a new trap according the distribution $\rho (\Gamma_m)$. Fig.
\ref{fig:landscape} shows a schematic of this dynamics of the mesoscopic region in the landscape. 

The mesoscopic regions are assumed to be large enough to be treated as deforming elastically until a ``plastic'' yield
event occurs.  In the trap picture, these plastic events are equated with the mescoscopic region leaving a trap.
The grains are driven by a constant shear rate in the experiments (Hartley \& Behringer (2003)). In a mean-field
approach similar to SGR, we assume this applied shear is transmitted without decay throughout the entire material. Thus
each mesoscopic
region is driven by the same shear rate.  Since the  mesoscopic regions deform elastically, the stress $\Gamma$ increases
linearly with time up to the point of
yielding:
 \begin{equation}
 \sigma =\frac{1}{S} \Gamma = k l =k \dot{\gamma }   t 
 \label{shearLINEAR}
 \end{equation}
where $\sigma$ is the shear stress, $S$ is the area of the mesoscopic region, and $k$ is the elastic constant relating the
shear stress to the shear strain, $l$.   The continual process of stress buildup and yielding creates a
``sawtooth" pattern in time as show in the bottom of Fig. \ref{fig:landscape}.

With a linear dependence of $\Gamma$ on time,  between yield events, we can rewrite Eq. \ref{sgrEOM}:
\begin{equation}
 \frac{
\partial }{
\partial  t}P \left(\Gamma _m,\Gamma , t\right)=- S k \dot{\gamma }  \frac{\partial }{
\partial  \Gamma }P \left(\Gamma _m,\Gamma , t\right)-\omega _0e^{-\alpha  \left(\Gamma _m-\Gamma \right)}P \left(\Gamma
_m,\Gamma , t\right)+\omega (t) \rho \left(\Gamma _m\right) \delta (\Gamma ) 
\label{sgr}
\end{equation}
The full time-differential on the l.h.s of Eq. \ref{sgrEOM} has been converted to a partial time derivative plus an
advective term. The delta function in the last term on the r.h.s. of Eq. \ref{sgr} means that immediatly after choosing
a new trap, the mesoscopic region is in a stress-free state. See points ``3" and ``5" in Fig. \ref{fig:landscape}.  If $\dot \gamma
=0$, the model reduces to Bouchaud's trap model in stress space. The steady state distribution,
$P(\Gamma_{m},\Gamma) \propto e^{-\alpha (\Gamma_m-\Gamma)} \rho(\Gamma_m) \delta(\Gamma)$, which is the equilibrium probability of finding a region with yield stress $\Gamma_{m}$. Since there is no imposed shearing, the instantaneous stress is zero.


Physically, hopping out of a trap equates to the grains in the mesoscopic region making a re-arrangement.
These events are visible in experiments and if analysed with photo-elastic techniques, are just buckling or the breaking
of force chains. After this re-arrangement, new force chains form. The newly formed force chains are characterized
by a new yield stress $\Gamma_{m}$.

The distribution of trap depths $\rho (\Gamma_m)$ describes the disorder in the stress landscape. Theoretically, this
distribution can be measured in experiments where no activated stress yielding occurs, or when stress fluctuations exist
only locally which leads to the angoricity ($1/\alpha$) being very small. This is the case when no neighboring stress fluctuation is ``felt"
 within the mesoscopic region. 
This is realized experimentally (Peidong \& Behringer (2009)). In this experiment, shear is applied to a small local
region which can be considered a single mesoscopic region. The stress yielding is due entirely to the driving. The
distribution of yield stresses in this case is exactly $\rho (\Gamma_m)$. Generally, in sheared granular systems (e.g.
Geng \& Behringer (2005)), we can get an idea for the form of $\rho (\Gamma_m)$ by looking at the tail of the stress
distribution (large $\Gamma$). This is because large $\Gamma$ values are accessed only when an activated process has not
occurred to make the the system yield. They are also states that sample large $\Gamma_m$ values. 

Activated
dynamics in stress space is not an entirely new concept.  It was proposed by Eyring (Glasstone {\it et al} (1941) ), and has been applied
to analyze the velocity profile in dense granular flows (Pouliquen \& Gutfraind (1996)).   In our work, the stress-activation is a
natural consequence of the ensemble based on angoricity.

We can write down a general solution to  Eq. \ref{sgr} regardless of the distribution of trap depth $\rho
(\Gamma_m)$.  Eq. \ref{sgr} can be simplified with a change of variable: $\delta \Gamma =\Gamma (t)- S k \dot{\gamma } t$,
which does not explicitly depend on time (away from the points of yielding). This eliminates the advective term in Eq.
\ref{sgr} which then becomes:
\begin{equation}
\frac{
\partial }{
\partial  t}P \left(\Gamma _m, \delta \Gamma , t\right)=-\omega _0e^{-\alpha \left(\Gamma _m-\left(\delta \Gamma +S k
\dot{\gamma }
t\right)\right)}P \left(\Gamma _m, \delta \Gamma , t\right)+\omega (t) \rho \left(\Gamma _m\right) \delta \left(\delta
\Gamma +S k
\dot{\gamma } t\right)
\label{sgrSUPER}
\end{equation}
 A better intuition can be gained by solving for  \ref{sgrSUPER} while ignoring the second term on its r.h.s., the
solution is:
\begin{equation}
 P\left(\Gamma _m, \delta \Gamma , t\right)=P_0\left(\Gamma _m, \delta \Gamma \right) \text{exp}\left(-\frac{\omega _0}{e^{\alpha 
\Gamma _m}}\int _0^tdt^{\prime} e^{\alpha  \left(\delta \Gamma +S k \dot{\gamma } t^{\prime}\right)}\right) 
\label{sgrDECAY}
\end{equation}
The solution is just an exponential decay with the time interval replaced by a time integral. This is the "effective time
interval" defined in (Sollich (1998)): 
 \begin{equation}
Z(t,t';\delta \Gamma )\equiv \int _{t'}^t dt'' e^{\alpha  \left(\delta \Gamma +S k \dot{\gamma } t''\right)}
\label{effetiveTIME}
\end{equation}
 In essence the behavior of an element in a trap is just that of a decaying process, $\sim e^{-t/\tau }$. For us, the
time is replaced by the effective time interval $Z(t, 0; \delta \Gamma )$ that grows much faster than linearly and the mean lifetime is $\tau =\omega
_0{}^{-1}e^{\alpha  \Gamma _m}$. After the stress collapse of the first buildup occurring at $t'$, the element escapes into
another trap chosen from the distribution $\rho (\Gamma _m)$ and starts to undergo its own decay process via the
effective time. Therefore the full solution is:

\begin{align} P\left(\Gamma _m, \delta \Gamma , t\right) =& P_0\left(\Gamma _m , \delta \Gamma \right) \text{exp}\left(-\frac{\omega _0}{e^{\alpha 
\Gamma _m}}Z(t,0 ; \delta \Gamma )\right) \\
& +\int _0^tdt'\omega (t') \rho \left(\Gamma _m\right)\delta \left(\delta \Gamma +S k \dot{\gamma }
t'\right)\text{exp}\left(-\frac{\omega _0}{e^{\alpha  \Gamma _m}}Z\left(t,t';\delta \Gamma \right)\right) \nonumber
\label{sgr_soln_full_int}
\end{align} 

\section{Theory predictions and comparison to experiments} 
Slow shearing experiments correspond to the long-time or steady state
limit of our model. In such a limit we find a solution that is time independent: 
\begin{equation}
P \left(\Gamma _m,\Gamma \right)= \rho \left(\Gamma _m\right)\text{exp}\left(-\frac{\omega _0}{S k \dot{\gamma } \;
\alpha \; e^{\alpha  \Gamma _m}}\left(e^{\alpha  \Gamma }-1\right)\right),
\label{sgr_soln}
\end{equation}
subject to normalization. We now have a ensemble distribution function of $\Gamma$ with which we can calculate the constitutive relation by taking
the ensemble average of $\Gamma$. Using Eq. \ref{shearLINEAR} we can write:
 \begin{equation}
 \sigma =\frac{1}{S} \langle \Gamma \rangle _{P\left(\Gamma _m, \Gamma \right)}=\frac{1}{S}\int _0^{\infty
}d\Gamma \; \Gamma \int _0^{\infty }d\Gamma _m \;   P \left(\Gamma _m,\Gamma \right)
 \label{constitutive}
 \end{equation}
We first consider the simplest case of a distribution of trap depths, $\rho (\Gamma_m) = \delta (\Gamma_m-\Gamma_0)$ with
$\Gamma_0 > 0$. Although not a physical description, this choice will prove to be a fruitful exercise. Eq.
\ref{constitutive} becomes:
 \begin{equation}
 \sigma =\frac{1}{\alpha  \; S}\frac{\int _1^{\infty }dy \; y^{-1} \log(y) e^{-A y}}{E_1(A)}
 \label{constitutive_delta}
 \end{equation}
 where the constant $A=\frac{\omega _0}{S k \dot{\gamma} \; \alpha }e^{- \alpha  \Gamma _0}$ and $E_1$ is the exponential
integral with $n=1$. In the interesting limit of $A \to 0$ or $ \dot{\gamma }>>\omega _0\left.e^{- \alpha  \Gamma
_0}\right/S k \alpha$, we can expand Eq. \ref{constitutive_delta} to $\mathcal{O}({A^0})$ and obtain:
 \begin{equation}
 \sigma =\frac{1}{2\alpha   S } \log\left(\dot{\gamma }\right)+\frac{1}{2\alpha  S } \left(\log\left( S k \alpha \left/\omega
_0\right.\right)+\alpha  \Gamma _0-2\gamma _c \right)
 \label{constitutive_delta_zeroth_log}
 \end{equation}
where $\gamma_c$ is the Euler-Mascheroni constant $\simeq 0.577$. The leading behavior of the constitutive equation has
a logarithmic dependence on the shear rate. This agrees qualitatively with experiments (Behringer {\it et al} (2008), Hartley \& Behringer (2003)).
In the opposite limit  $A \to \infty$ or $\dot{\gamma } << \omega _0\left.e^{- \alpha  \Gamma_0}\right/S k \alpha$, we can expand Eq. \ref{constitutive_delta} to $\mathcal{O}({1/A})$ and obtain:
 \begin{equation}
 \sigma =\frac{k \; e^{ \alpha  \Gamma _0}}{\omega_0  } \dot{\gamma}
 \label{constitutive_delta_zeroth_Newtonian}
 \end{equation}
 This reveals a Newtonian regime with viscosity $\eta={k \; e^{ \alpha  \Gamma _0}}/{\omega_0  }$. 
 
 \subsection{Constitutive equation for the exponentially decaying distribution of trap depths}
\begin{figure}[htp]
\centering
\includegraphics[width=4in]{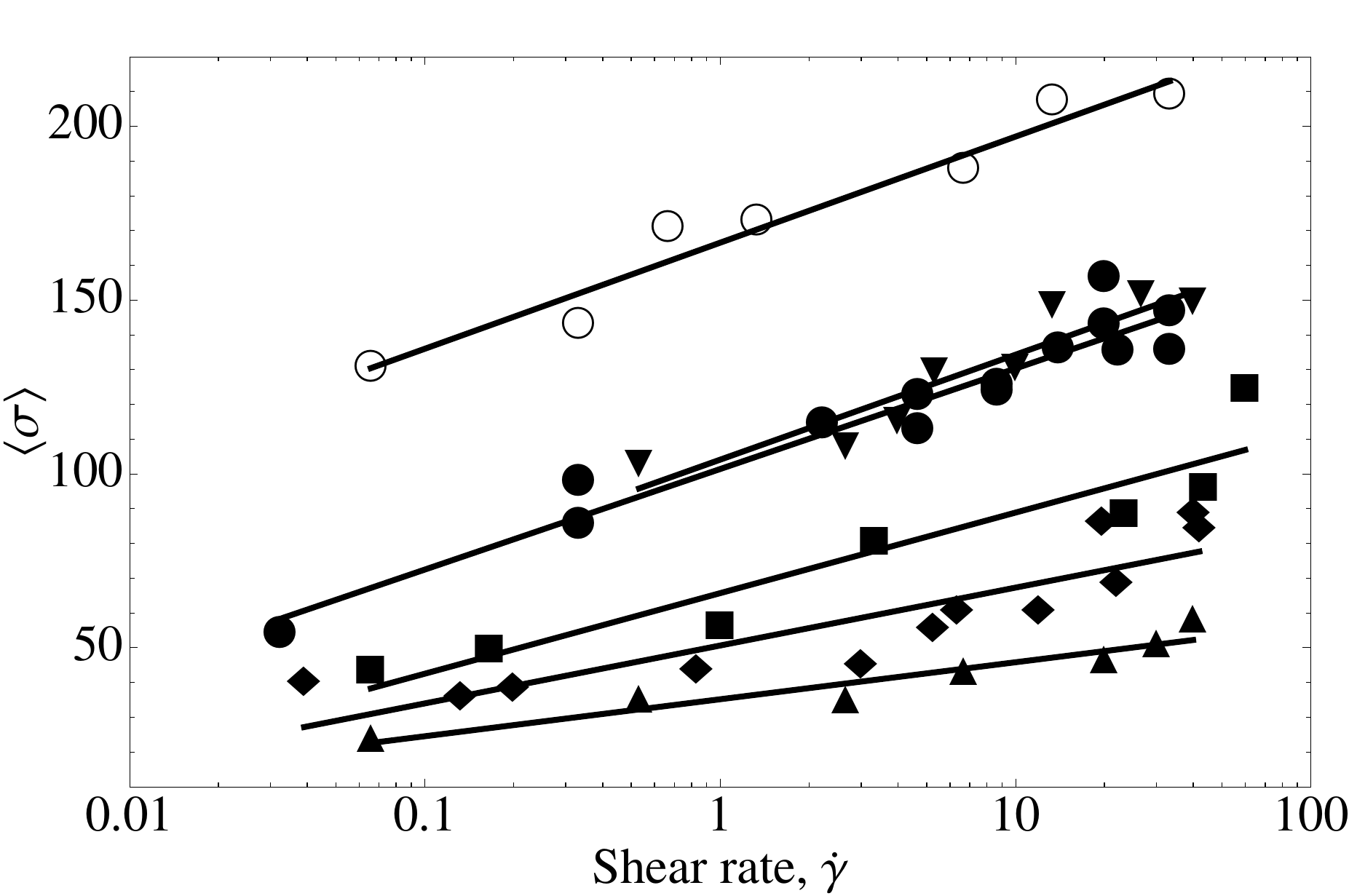}
\caption{Linear-Log Plot of mean stress in a segment of 2D granular Couette experiment containing roughly 200 particles
vs. shear rate. Data, taken from (Hartley \& Behringer (2003)), are given for different packing fractions, which are
given relative to the critical packing
fraction $\phi_c$ where the system loses all rigidity. Various symbols are: upward triangles: $\phi=\phi_c+0.00035$;
diamonds:
$\phi=\phi_c+0.00141$; squares: $\phi=\phi_c+0.00528$; solid disks: $\phi=\phi_c+0.00528$; downward triangles:
$\phi=\phi_c+0.00800$; and hollow circles: $\phi=\phi_c+0.01373$.  }
\label{fig:constitutive}
\end{figure}

 As discussed in the previous section, an exponentially decaying $\rho (\Gamma_m)$ is a better description of granular experiments as well as spin-glasses. We use:
 \begin{equation}
 \rho (\Gamma_m) =  {1 \over \alpha_{0}} e^{- \alpha_0 \Gamma_m}
 \label{exp_dos}
 \end{equation}
where $\alpha_0$ is a constant measuring the disorder in the stress landscape. Inserting Eq. \ref{exp_dos} into Eq. \ref{constitutive} we obtain:
 \begin{equation}
 \sigma =
 \frac{1}{\alpha  S }\log(S k \dot{\gamma} \; \alpha / \omega _0)+
 \frac{1}{\alpha S }
 \frac	{\int _0^{\infty }\frac{dW}{W^x} \frac{\log(W+\omega _0 / S k \dot{\gamma}\alpha)} 
 		{(W+\omega _0 / S k \dot{\gamma}\alpha)}
 \gamma \left(x ,W\right)}
{\int _0^{\infty }\frac{dW}{W^x (W+\omega _0 / S k \dot{\gamma}\alpha
)} \gamma \left(x ,W\right)}
 \label{constitutive_exp}
 \end{equation}
 where x is given by the ratio: $x=\alpha_0 / \alpha$, and $\gamma (x, W)$ is the lower incomplete gamma function. Again we look in the limit of 
$ \dot{\gamma }>>\omega _0\left. \right/k \alpha$ and the constitutive equation is approximated to zeroth order as:
 \begin{equation}
\sigma=\frac{2}{\alpha S  } \log( \dot{\gamma})+\frac{1}{\alpha S }\log (\frac{S k \alpha ^2}{\omega _0\alpha _0} )
 \label{constitutive_exp_zeorth}
 \end{equation}
 This shows that the logarithmic dependence of shear stress on shear rate is independent of the form of the distribution $\rho
(\Gamma_m)$ used. In Fig. \ref{fig:constitutive} we compare this result with the data obtained in 2D Couette granular shearing experiments
(Hartley \& Behringer (2006)). 

In the opposite limit, $ \dot{\gamma }<<\omega _0\left. \right/S k \alpha$ we again obtain a Newtonian regime:
\begin{equation}
\sigma=\frac {k}{\omega_0}\frac{x-1}{x-2} \; \dot{\gamma}
\label{constitutive_exp_zeorth_newtonian}
\end{equation}
The viscosity diverges at $x=2$, the same result is obtained in SGR (Sollich (1998)).

\subsection{Yield Stress}
\begin{figure}[htp]
\centering
\includegraphics[width=4in]{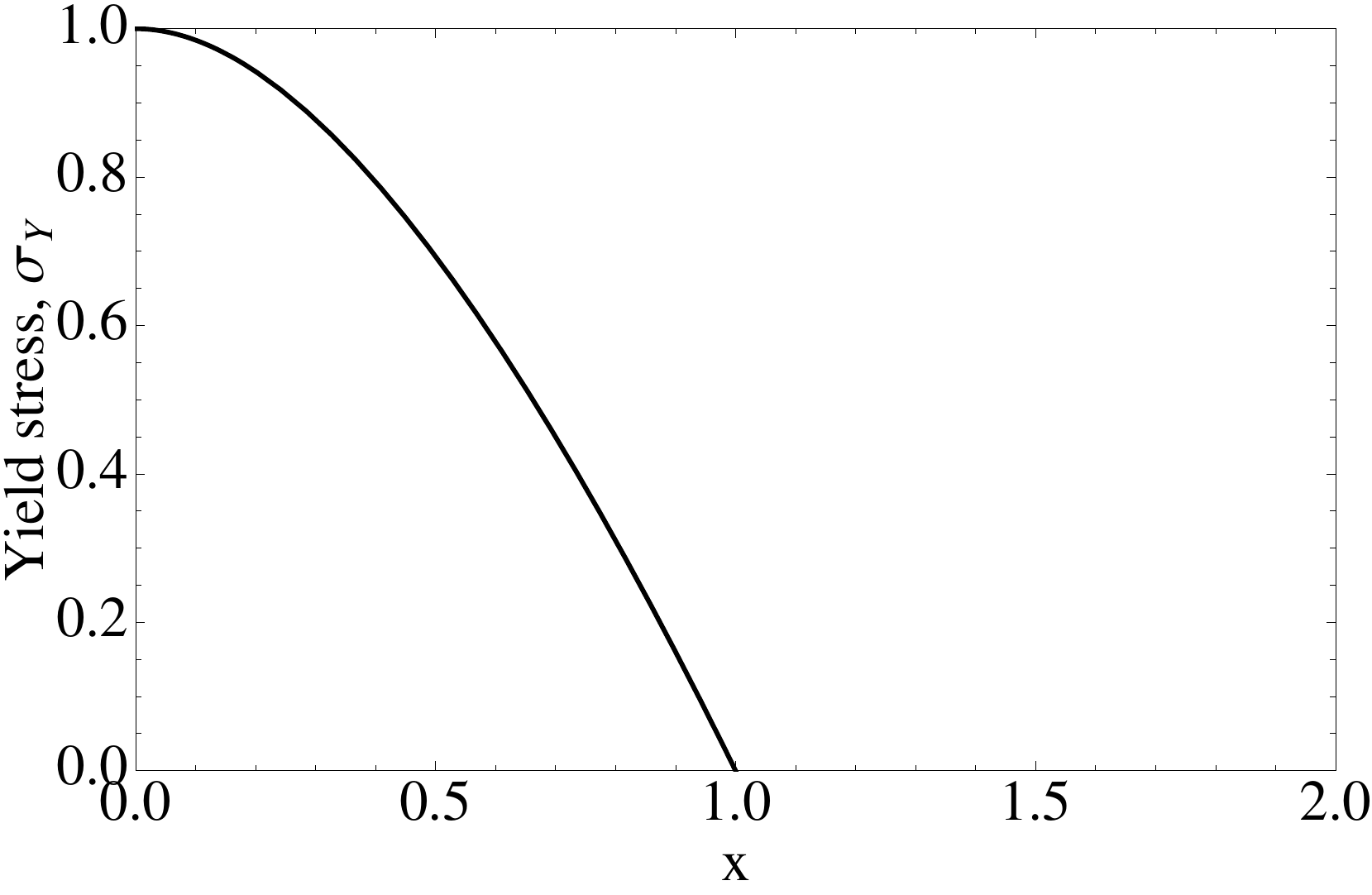}
\caption{Theoretical Plot of the yield stress Eq. \ref{yield_stress} as a function of $x=\alpha_0 / \alpha$. It vanishes at x=1. } 
\label{fig:yield_stress} 
\end{figure}

When $ \dot{\gamma}=0$, the stress has a finite value for $\alpha_0<\alpha $ or $x<1$. This is
the yield stress,  and it vanishes at the same point as the glassy transition point in Bouchaud's trap model. The
yield stress can be calculated by inserting $\dot{\gamma}=0$ into Eq. \ref{constitutive}. Performing the $\Gamma$ integral first in Eq. \ref{constitutive}, we obtain:
 \begin{equation}
 \sigma = \frac{1}{\alpha S} \frac {\int _0^{\omega_0/S k \dot{\gamma} \; \alpha}dz \; z^{x-1} G(z)} {\int _0^{\omega_0/S k \dot{\gamma} \; \alpha}dz \; z^{x-1} E_1(z)},
 \label{yield_stress_0}
 \end{equation}
where $G(z)$ is the Meijer G-function:
 \begin{equation}
G(z|\begin{array}{c}1, 1 \\ 0,0,0\end{array})= \int_1^{\infty} dy \frac{\rm{log}(y)}{y} \rm{exp}(-z (y-1))
\nonumber
 \end{equation}
 When $\dot{\gamma}=0$, Eq. \ref{yield_stress_0} can be calculated exactly:
 \begin{equation}
 \sigma_y \equiv \sigma ( \dot{\gamma } = 0)= -k \; x \; \gamma_c-k \; x \frac{d}{d x} \log\left(\Gamma(x)\right) ,
 \label{yield_stress}
 \end{equation}
A plot of the
yield stress is shown in Fig. \ref{fig:yield_stress}

\subsection{The distribution of stress drops} 
It is often easier to measure distribution of stress drops in experiments than the stresses themselves, and the
distribution provides a more stringent test of theoretical frameworks.
While in a trap with depth $\Gamma_m$, the probability to build up the stress
to a value of $\Gamma$ is given by: ({\it cf} Eq. \ref{sgr_soln} )
\begin{equation}
P_s(\Gamma ) \propto exp \left(-\frac{\omega _0}{S k \; \dot{\gamma } \; \alpha \; e^{\alpha  \Gamma _m}}\left(e^{\alpha  \Gamma
}-1\right)\right) \nonumber
\label{prob_s}
\end{equation}
On the other hand, as the stress grows, it is increasingly likely to fail due to the activated process. The rate of failure is given by:
\begin{equation}
P_f(\Gamma ) \propto e^{-\alpha  \left(\Gamma _m-\Gamma \right)} \nonumber
\label{prob_f}
\end{equation}
Then, the probability of building the stress up to $\Gamma$ and then to fail at this point is given by the product $P_s \times  P_f$. Since the stress falls to zero after a failure event, we can call the magnitude of the stress drop $\Delta \Gamma$. Its distribution for a given $\rho(\Gamma_m)$ is proportional to:
\begin{equation}
 P(\Delta \Gamma )\propto P_s(\Delta \Gamma )P_f(\Delta \Gamma )
 \nonumber 
 \end{equation} 
Also sum over all possible traps:
 \begin{equation}
P(\Delta \Gamma ) \propto \int d\Gamma _m\rho \left(\Gamma _m\right)P_s(\Delta \Gamma )P_f(\Delta \Gamma )
 \end{equation} 
 Using the exponentially decaying distribution $\rho (\Gamma_m) = e^{- \alpha_0 \Gamma_m}$, we obtain the normalized distribution:
 \begin{equation}
P(\Delta\Gamma)=x \; \alpha \left(\frac{\omega_0}{S k \dot{\gamma} \alpha}\right)^{-x} e^{\alpha \Delta\Gamma} 
 \left(e^{\alpha \Delta\Gamma}-1 \right)^{-1-x} \; \gamma \left(1+x, \; \frac{\omega_0}{S k \dot{\gamma} \alpha} (e^{\alpha \Delta\Gamma}-1 ) \right)
\label{stress_drop_distribution}
\end{equation}
where we see the lower incomplete gamma function again. There are three regimes in which Eq. \ref{stress_drop_distribution}
can be simplified. First, using the limit $y^{-\beta} \; \gamma(\beta,y) \to 1/ \beta$ as $y \to 0$, we get:
\begin{equation}
P(\Delta\Gamma)=\left(  \frac{x}{x+1} \right)\frac{\omega_0}{S k \; \dot{\gamma}} \; e^{\alpha \Delta\Gamma}  \;\;
when \hspace{10 pt} \frac{\omega_0}{S k \dot{\gamma} \alpha} (e^{\alpha \Delta\Gamma}-1 ) \to 0
\label{stress_drop_distribution_case1}
\end{equation}
The opposite limit for the lower incomplete gamma function is $\gamma(\beta,y) \to (\beta - 1)!$ as $y \to \infty$ for which we get:
\begin{equation}
P(\Delta\Gamma)=\alpha \; x^2 (x-1)! \left(\frac{\omega_0}{S k \; \dot{\gamma} \alpha}  \right)^{-x} e^{\alpha \Delta\Gamma} 
\left( e^{\alpha \Delta\Gamma} -1 \right)^{-1-x}
when \hspace{10 pt} \frac{\omega_0}{S k \dot{\gamma} \alpha} (e^{\alpha \Delta\Gamma}-1 ) \to \infty
\label{stress_drop_distribution_case2}
\end{equation}
Finally, the more extreme case of Eq. \ref{stress_drop_distribution_case2} as $\Delta\Gamma \to \infty$:
\begin{equation}
P(\Delta\Gamma)=\alpha \;  x \;x! \left(\frac{\omega_0}{S k \; \dot{\gamma} \alpha}\right)^{-x} e^{-\alpha_0 \Delta\Gamma} 
when \hspace{10 pt} \Delta\Gamma \to \infty
\label{stress_drop_distribution_case3}
\end{equation}
In Eq. \ref{stress_drop_distribution_case3}, we recover the exponential distribution of depths. 

\begin{table} 
\centering
\label{fit_results}
\caption{Results of fitting Eq. \ref{stress_drop_distribution_case2} to the experimental data of Hartley \& Behringer
(2003)}
\begin{tabular}{lllll}
\hline
$\phi - \phi_c$  & 
Strain Rate   $\dot{\gamma} $ (mHz)& 
$\alpha$ & $\alpha_0$ & 
$x=\alpha_0/\alpha$ \\
\hline
\hline

0.0053	&	0.3318	&	0.7000	&	0.32473	&	0.4639	\\
		&	2.2148	&	1.0000	&	0.2339	&	0.2339	\\
		&	4.6568	&	1.4000	&	0.19054	&	0.1361	\\
		&	8.6491	&	2.0000	&	0.1624	&	0.0812	\\
		&	13.9722	&	4.0000	&	0.1244	&	0.0311	\\
\hline
0.0014	&	0.0390	&	0.6000	&	0.34872	&	0.5812	\\
		&	0.1322	&	0.6000	&	0.3279	&	0.5465	\\
		&	0.1987	&	0.6000	&	0.34566	&	0.5761	\\
		&	0.8308	&	0.7000	&	0.3115	&	0.4450	\\
		&	2.9933	&	0.9000	&	0.25857	&	0.2873	\\
		&	5.2490	&	1.0000	&	0.2366	&	0.2366	\\
		&	6.3203	&	1.2000	&	0.21372	&	0.1781	\\
		&	11.9760	&	1.3000	&	0.18083	&	0.1391	\\
		&	19.6279	&	4.0000	&	0.1428	&	0.0357	\\
		&	21.9567	&	6.0000	&	0.1836	&	0.0306	\\
		&	0.0657	&	0.5000	&	0.30135	&	0.6027	\\
		&	0.5314	&	0.5000	&	0.26265	&	0.5253	\\
		&	2.6607	&	0.7000	&	0.24444	&	0.3492	\\
		&	6.6529	&	0.8000	&	0.1928	&	0.2410	\\
		&	11.9760	&	1.1000	&	0.17061	&	0.1551	\\
		&	19.9606	&	1.2000	&	0.17316	&	0.1443	\\
		&	29.9413	&	1.2000	&	0.15372	&	0.1281	\\
		&	39.9221	&	1.5000	&	0.1344	&	0.0896	\\
\hline
0.0088	&	0.5314	&	0.5000	&	0.2375	&	0.4750	\\
		&	2.6607	&	0.9000	&	0.25146	&	0.2794	\\
		&	3.9848	&	1.0000	&	0.2282	&	0.2282	\\
		&	5.3155	&	1.2000	&	0.21312	&	0.1776	\\
		&	9.9998	&	1.6000	&	0.1776	&	0.1110	\\
		&	13.3068	&	2.5000	&	0.163	&	0.0652	\\
\hline
0.0137	&	0.0657	&	0.4000	&	0.25472	&	0.6368	\\
		&	0.3318	&	0.5000	&	0.33945	&	0.6789	\\
		&	0.6645	&	0.5000	&	0.27425	&	0.5485	\\
		&	1.3299	&	0.5000	&	0.26505	&	0.5301	\\
		&	2.6607	&	1.3000	&	0.33384	&	0.2568	\\
		&	3.3260	&	1.4000	&	0.24458	&	0.1747	\\
		&	6.6529	&	1.5000	&	0.2097	&	0.1398	\\
		&	13.3068	&	4.0000	&	0.1776	&	0.0444	\\

\end{tabular}

\end{table}

\begin{figure} [htp]
  \begin{center}
    \subfigure[]{\label{fig:fit-a} \includegraphics[width=2.25in]{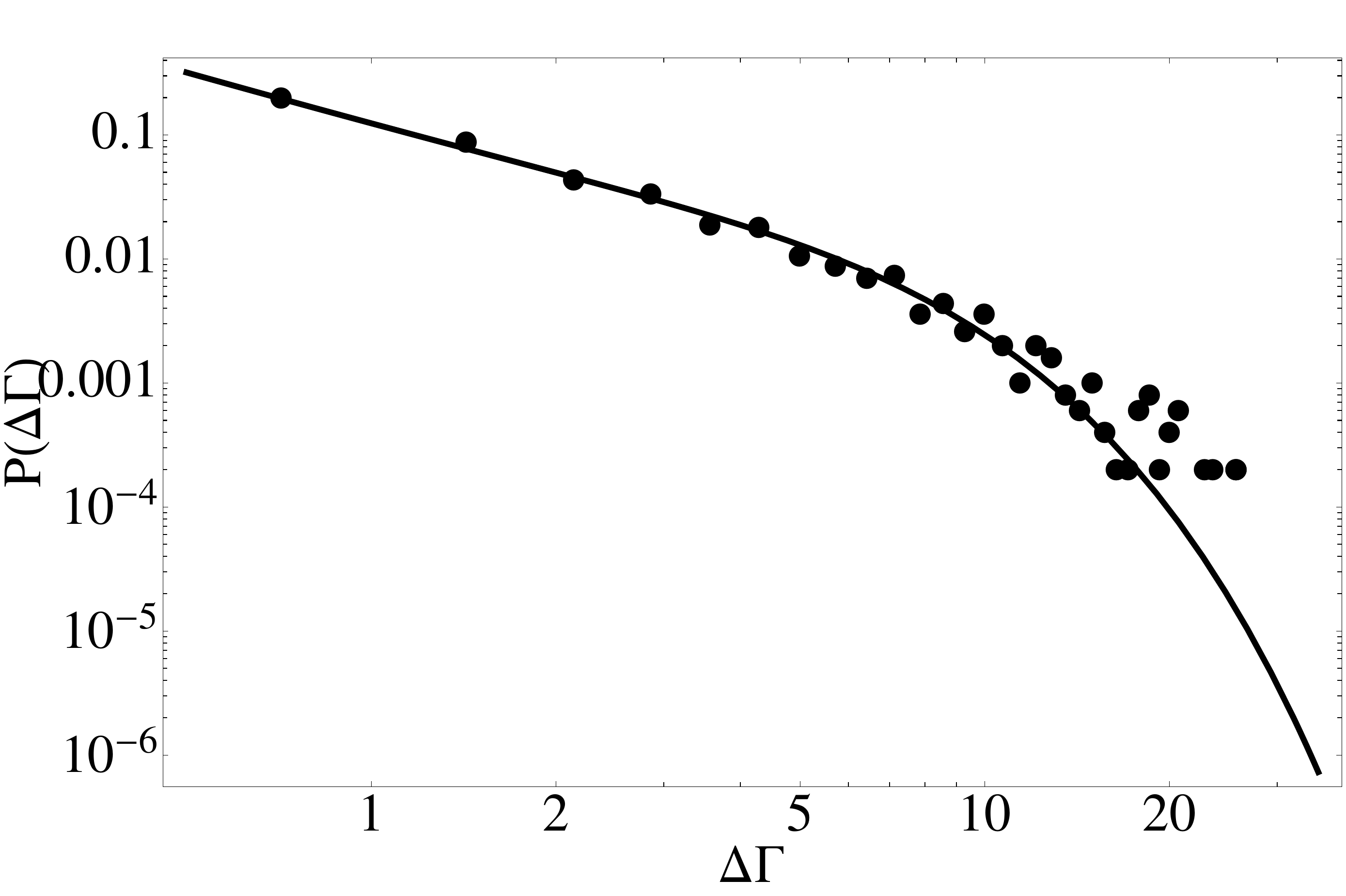}} 
    \subfigure[]{\label{fig:fit-b} \includegraphics[width=2.25in]{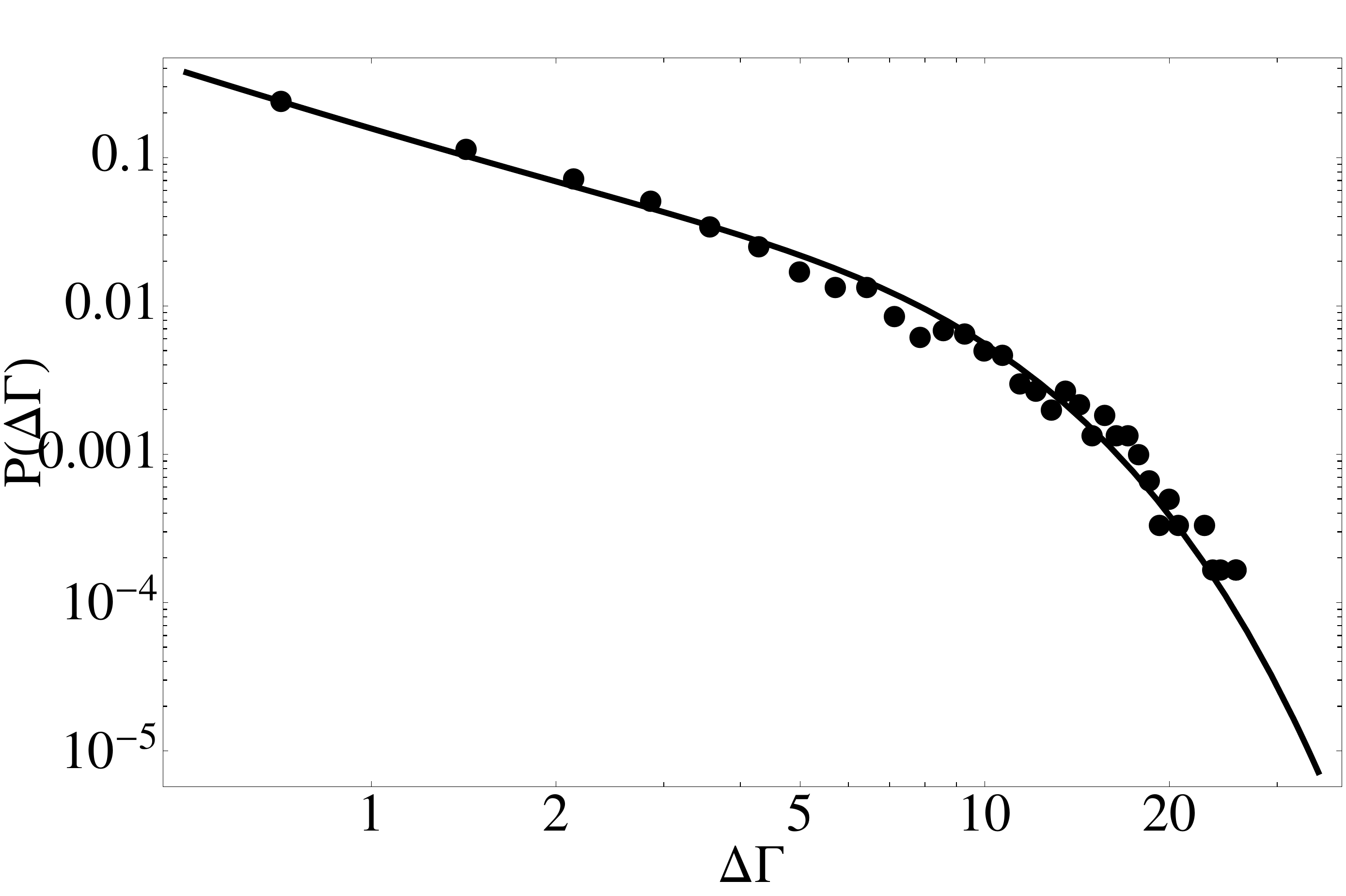}} \\
    \subfigure[]{\label{fig:fit-c} \includegraphics[width=2.25in]{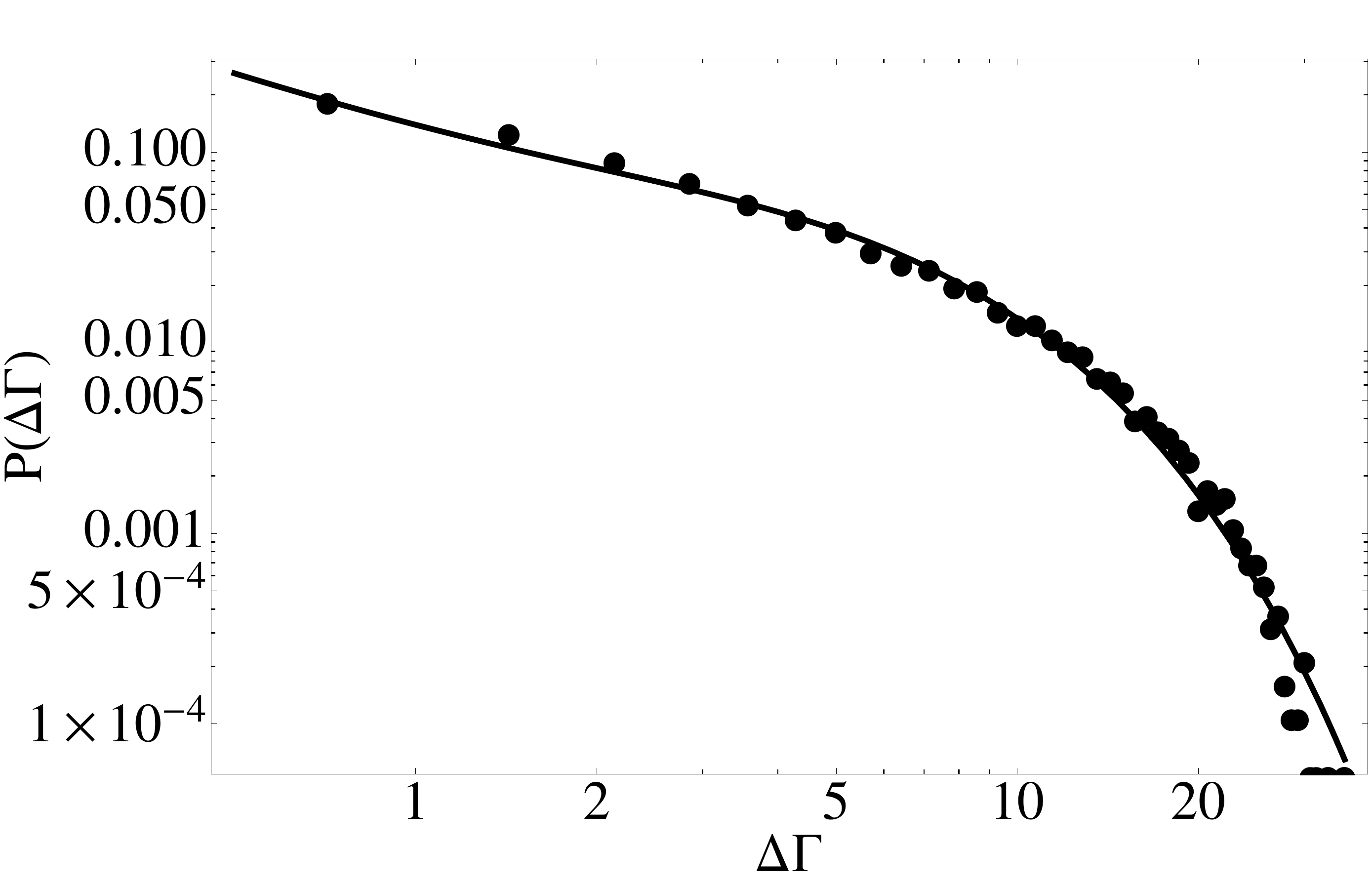}} 
    \subfigure[]{\label{fig:fit-d} \includegraphics[width=2.25in]{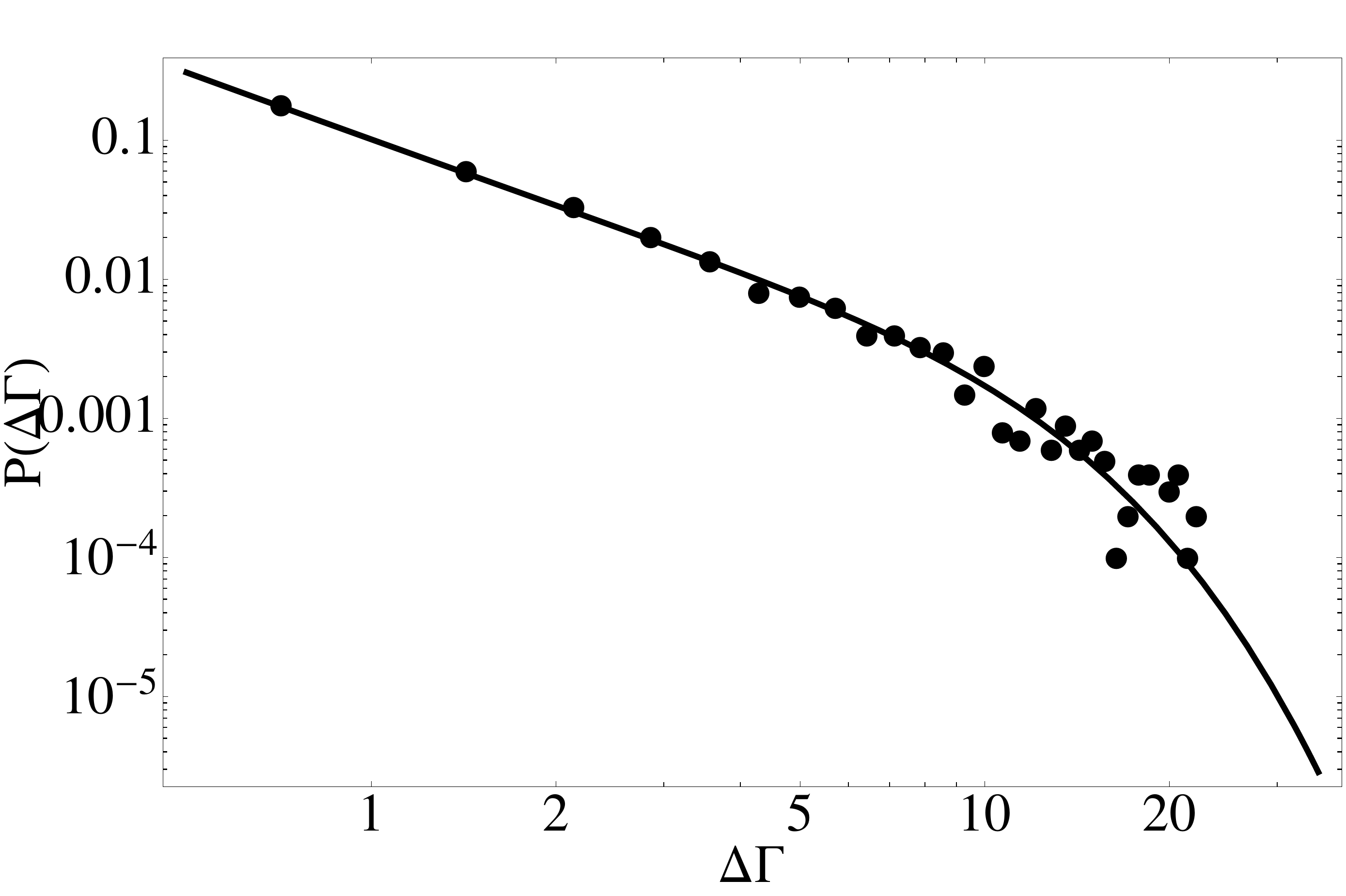}}   
  \end{center}
  \caption{Log-Log plot of experimental distribution of stress drops and its fitting to Eq. \ref{stress_drop_distribution_case2}. The experimental and fitting 	parameters are: 
  (a) $\dot{\gamma}=0.3318mHz, $ $ \phi-\phi_c=0.0053, $ $ \alpha=0.70,$ $ \alpha_0=0.3247$, 
  (b) $\dot{\gamma}=0.9972mHz, $ $ \phi-\phi_c=0.0053, $ $ \alpha=0.7,$ $ \alpha_0=0.2660$ , 
  (c) $\dot{\gamma}=5.3155mHz, $ $ \phi-\phi_c=0.0088, $ $ \alpha=1.2,$ $ \alpha_0=0.2131$, and 
  (d) $\dot{\gamma}=0.0657mHz, $ $ \phi-\phi_c=0.0137, $ $ \alpha=0.4,$ $ \alpha_0=0.2547$.}
\label{fig:fit}
\end{figure}

\begin{figure}[htp]
\centering
\includegraphics[width=4in]{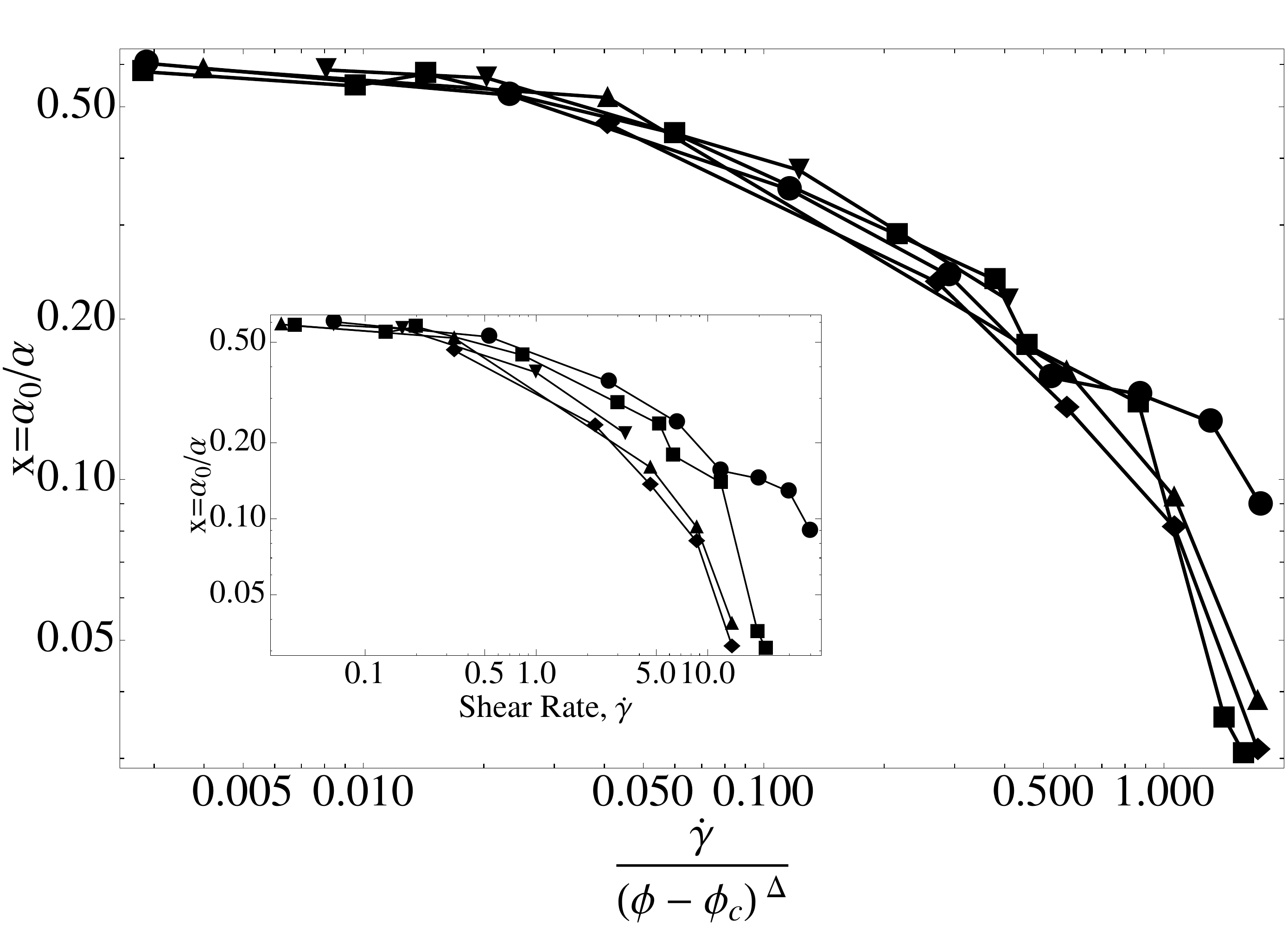}
\caption{Log-log plot of the scaled the relationship (Eq. \ref{xSCALED}). The scaling exponent is found to be: $\Delta
=-0.4$ The data points are: disks: $\phi-\phi_c=0.0004$, squares: $\phi-\phi_c=0.0014$, diamonds: $\phi-\phi_c=0.0053$,
upward triangles: $\phi-\phi_c=0.0053$, and downward triangles: $\phi-\phi_c=0.0053$. \newline \textbf{Inset:} Unscaled
log-log plot of $x$ vs. $\dot {\gamma}$.  } 
\label{fig:xVSshearrate}
\end{figure}

With an analytical form of $P(\Delta \Gamma)$, we can compare and fit to data obtained by Hartley \& Behringer (2003).
The data is
fitted to the form Eq. \ref{stress_drop_distribution_case2} and therefore contains two fitting parameters: $\alpha$ and
$\alpha_0$. Table~1 shows the fitting results. In Fig. \ref{fig:fit}, four fitting results are plotted and compared with
data.   As seen from Table ~1, both $\alpha$ and $\alpha_{0}$ are functions of the shear rate and the packing
fraction. The
data fitting gives us insights into the relationship between the shear-angoricity ($x=\alpha_0 /
\alpha$), the shear
rate ($\dot {\gamma}$), and the packing fraction $\phi-\phi_c$. We can see this relation in a plot of $x$ vs. $\dot
{\gamma}$ in Fig. \ref{fig:xVSshearrate} for different $\phi-\phi_c$.  
The first observation one makes is that $x$ is controlled by both $\dot {\gamma}$, and  $\phi-\phi_c$,  implying that
the SGR assumption of $x$ being independent of the shear rate does not apply to these measurements in sheared,
granular media.  In addition, over a decade of $\phi$
-$\phi_{c}$ values, and three decades of $\dot \gamma$, $x$
seems to be described by a scaling form: 
\begin{equation}
x=f_+ \left( \frac{\dot {\gamma}}{{|\phi-\phi_c|}^\Delta} \right) ~,
\label{xSCALED}
\end{equation}
with $\Delta \simeq -0.4$, and the scaling function (for $\phi \ge \phi_{c}$) $f_{+} (z)$ being a decreasing function
of its argument, $z$.  The dependence on $\dot \gamma$ is, however, weaker than the dependence on $\phi - \phi_{c}$.
Although better
statistics are needed to pin down the scaling form, it is intriguing to examine its consequences as such.   The scaling
itself hints at $(\phi-\phi_{c}=0, \dot \gamma =0)$ being a critical point, which is consistent with the observations
of Olsson \& Teitel (2007), and the idea of Point J being a special point in the jamming phase diagram (Liu \& Nagel
(1998)). At zero shear rate, $x$ becomes independent of $\phi - \phi_{c}$, and
the scaling form is, therefore, consistent with the packings possessing a yield stress for all $\phi > \phi_{c}$.   
From the perspective of $x$
being the fluctuation temperature, the observation is that stress fluctuations grow as $\phi$ approaches $\phi_{c}$ from
above.  Experiments do indicate growing fluctuations as a system approaches the unjamming transition below which it
cannot sustain shear (Howell \& Behringer (1999).

\section{Conclusion} 
A stress-based statistical ensemble has been used in conjunction with the idea of a stress landscape to construct a
model for granular rheology.  Comparisons of the model to experiments on slowly sheared granular media 
indicate that the theoretical framework provides  a semiquantitative description of the stress-response.   The
framework is an extension of the Soft Glassy Rheology approach, and the stress ensemble provides a natural definition
for the fluctuation temperature.  This fluctuation temperature can be obtained by fitting the theoretical predictions
to experiments, and the results show that fluctuations are largest at the special point, Point J, where the shear rate
goes to zero and the packing fraction approaches the critical value below which one cannot construct a mechanically
stable state of grains.  In future work, we intend to extend the model to include volume fluctuations since Reynold's
dilatancy is one of the signature properties of granular matter (Reynold (1885)).  A fundamental premise of the stress
ensemble framework is that just as in thermal equilibrium temperature is equalized, for a system in granular
equilibrium, the angoricity is the same everywhere inside the system. Comparison to simulations has shown this to be
true for a system under pure compression (Henkes {\it et al} (2007)).  If we assume the same to be true for the
shear-angoricity $x$, then according to Eq. \ref {xSCALED},  the local packing fraction has to adjust to the
local shear rate and regions of increased shear rate have lower packing fractions.
An interesting question to ask is
whether this is the origin of Reynold's dilatancy.
\begin{acknowledgements}
We acknowledge many helpful discussions with Mike Cates, R. Behringer, Trush Majmudar, Gregg Lois, Jie Zhang, Mitch Mailman, and Corey
O'Hern. 
DB acknowledges the hospitality of Duke University where some of this work was done, and BC acknowledges the
hospitality of the Aspen Center for Physics, where a lot of the ideas took shape.  This work was funded by the grant NSF-DMR
0549762.
\end{acknowledgements}

\end{document}